\documentclass[11pt]{article}
\usepackage{graphicx}
\usepackage{latexsym}

\begin{document}

\title{Proper Mass Variation under Gravitational and Coulomb Force Action in Relativistic Mechanics of Point Particle}
\author{Anatoli Andrei Vankov\\         
{\small \it IPPE, Obninsk, Russia; Bethany College, KS;  anatolivankov@hotmail.com}}

\date{}

\maketitle

\begin{abstract}

The problem studied is formulated in the title: proper mass variation under gravitational and Coulomb force action in Relativistic Mechanics of point particle. The novelty is that equations of motion are obtained in the relativistic Lagrangean framework for conservative force fields under assumption of field dependent proper mass. The dependence of proper mass on field strength is derived from the equations of particle motion. The result is the  elimination of a classical $1/r$ divergence. It is shown that a photon in a gravitational field may be described in terms of a refracting massless medium. This makes the gravity phenomenon compatible with SR Dynamic framework. New results concerning gravitational properties of particle and photon, as well as an experimental test of predicted deviation from $1/r^2$ classical Coulomb force law are discussed. The conclusion is made that the approach of field-dependent proper mass is perspective for better understanding GR problems and further studies on divergence-free field theory development.


\end{abstract}

\section{Introduction and Brief Overview}

{\em Objective, novelty, and framework}

The multi-aspect problem studied is the proper mass variation under gravitational and Coulomb force action in Special Relativistic Mechanics of point particle. 
We start with the Special Relativity question: is a gravitational force compatible with SR? When investigating it, we do not use arguments from a quantum field theory, and do not question General Relativity. The objective of this work is to show how a gravitational force along with the Coulomb force can be included in the SR Mechanics. 
Thus, the problem of relativistic mechanics of point particle in $1/r$ potential field was formulated in Lagrangian terms and studied in SR mechanics framework.  A novelty of our approach is that the proper mass varies under the force action, and its dependence on field strength is found from the Euler-Lagrange equations in SR-based Dynamics. 

Though we use the term ``field'', it has a classical mechanics meaning of $1/r$ potential field, or the corresponding Minkowski force field. We avoid the term ''scalar field'', which is used in field theories in analogy with ``vector and tensor field'', because those terms are irrelevant to SR Mechanics of point particle, we deal with.  Occasionally we refer to some comparable results of ``conventional theories'' as far as it concerns problems of particle motion discussed in conventional SR Mechanics as well as in GR or classical field theories under assumption of proper mass constancy.

{\em SR/gravity compatibility}

At some historical stage of GR development, there were numerous attempts to incorporate the Newton's formulation of the gravitational law into SR as a starting point to a field theory development. A Newtonian field propagates with infinite velocity, and one could expect that this assumption would be automatically corrected in the covariant formulation of the gravitational law. Approaches were based on the concept of proper mass constancy and the concept of a photon coupling to the gravitational field: the latter was thought a necessary condition for explaining the observed bending of light (see \cite{Misner} and elsewhere). Not surprisingly, the attempts failed, first of all, because the stress-energy tensor of the electromagnetic field has a vanishing trace. Thus, SR Mechanics of a point particle under gravitational force action has never been developed.

{\em Gravitational proper mass variation}

We revisited this problem in the SR framework and studied the role of the proper mass in SR dynamics. The conclusion was made that the commonly used concept of the proper mass constancy is neither required in theory physical foundations nor it is justified by observations: so far, this is an arbitrary assumption, subject to theoretical examination and experimental verification. 

In our SR-based methodology of a variable proper mass, the world line is curved, while the metric remains the Minkowskian one: diagonal elements are functions of dynamical variables, off-diagonal terms identically equal zero, while the proper time interval is not Lorentz invariant. Given the corresponding geometry, the equations of particle motion were derived, solutions to which were found similar to those in  GR (Schwarzschild field) dynamics {\it under weak-field conditions}. Differences rise with field strength.

{\em Radial motion in Coulomb field}

The problem of a point charge motion in the Coulomb macroscopic field is studied in the same methodology. It is assumed that the electric constant is field dependent consistently with the proper mass variation and the gravitational refraction concept so that the ratio of gravitational and Coulomb force is constant. As a result, equations of motion were obtained similar to that in the gravitational force case. A test of the predicted deviation from $1/r^2$ classical Coulomb force law in high voltage experiments is proposed. 

{\em Results}

New results were predicted concerning gravitational properties of a particle {\it under strong-field conditions}. One of the results is the elimination of the $1/r$ divergence in the solutions. As for photon, the conclusion was made that it can be treated in a relativistic model, in which a field acts on the photon as an optically active medium. In other words, this is the gravitational refraction rather then force attraction that causes the bending of light. Thus, the issue of SR incompatibility with the gravity phenomenon took a new turn: the inclusion of gravitational forces into SR was justified.


{\em Interfacing issues}. 

There are important problems relevant to SR Dynamics, which are out of the scope of the present work, for example, an interaction of two neutral or charged objects of comparable proper masses. There is an understanding of how those problems can be handled in the alternative SR dynamics. On the other hand, there are topics, which are traditionally ``difficult'' in field theories, such as a many-particle system, a radiation during acceleration, and the corresponding reaction force (in both the gravitational dynamics and electrodynamics). They are discussed in current literature (see, for example, \cite{Fields}). 
In our view, difficulties arising in field theories are caused by the divergence problem in field theories and quantum character of the phenomena in question. 

As was emphasized, the SR Dynamics problem studied in the present work has nothing to do with any field theory (a quantum field theory, in particlar). Nevertheless, we pay attention to the question of how our results could be useful for better understanding of the nature of the divergence problem in current field theories, in particular, GR. The matter is that GR is a next level of theory with respect to Minkowski force mechanics. A field theory (QFT first of all) is intended to specify and explain a mechanism of particle-particle (or particle-source) interaction via a field quantum of a certain spin (say, scalar, vector, tensor type of interaction). However, GR is not ``an ordinary'' field theory. Its peculiarity is two-fold. Firstly, it fails to follow conventional renormalization and quantization rules known to be effective in QED; secondly, concepts of force and kinetic/potential energy are abandoned in favor of a general curved spacetime concept.    

{\em Comments}

{\em Quantization}. GR non-quantizibility (Quantum Gravity problem) is one of the Physics Frontiers problem. In our view, an incorporation of the de Broglie wave concept under gravitational dynamical conditions into GR could be a first step to the problem solution. In this work, we discuss quantum connections arising in SR-based gravitational dynamics due to the de Broglie 4-wave introduction in the variable proper mass approach. Could quantum connections be established in a similar way in GR? 

{\em Divergence}. In our view, further studies are needed to find out if different types of divergence in classical field and quantum field theories have a common root, namely, the assumption of proper mass constancy. It should be noted that $1/r$ Schwarzschild divergence would be eliminated if the gravitational GR gauge factor $1/(1-2 r_g/r)$ were replaced by 
the factor $1/{\gamma_r}^2=\exp{-(2 r_g/r)}$. The latter naturally arises in the variable proper mass approach. This could make a difference in the gravitational radiation problem. Our idea of singularity elimination was presented earlier in (\cite{Vankov1}), and here we study different aspects of the problem in more details.

{\em Force action versus spacetime curvature}. The GR spacetime curvature concept looks like a radical deviation from concepts of force and kinetic/potential energy in classical and relativistic Physics. However, our study of Noether's conservative currents and corresponding dynamical  symmetries in complementary 4-coordinate and 4-momentum spaces with a gravitational source led us to the conclusion that both descriptions could be equivalent reflections of the same reality, at least, under weak-field conditions. Differences in our and GR predictions rise with field strength, obviously, because of the desired divergence elimination in our approach. Thus, the revision of the proper mass concept opens new opportunities in a gravitational theory development.

{\em Our central claim} 

The claim is the consistent incorporation of the gravitational force along with the Coulomb force into the SR-framework for the particular case of a point particle in the $1/r$-potential field in the approach of the field-dependent proper mass. New predictions under the strong-field conditions are obtained and discussed. We believe that the approach is perspective for further studies on divergence-free field theory development. 

{\em Readers' profile}. 

While discussing the problem formulation and obtained results, we bore in mind that the work could be interesting not only to specialists in the SR and GR areas; we expect that physicists, engineers, teachers and students from different branches of Physics and technology, who want to learn more about relativistic gravitational Physics, will find useful information. 

{\em Content}. 

The Lagrangean problem formulation and its solution are given in Sections 2, 3, Lagrangian symmetries and Noether's conservative currents related to the energy conservation law, and graphical illustrations are given in Section 4. Section 5 is devoted to the photon problem. Predictions and observations are discussed in Section 6. Section 7 is finalizing.

\section{Lagrangean Formulation of Relativistic Mechanics of Point Particle in Gravitational Field}

\subsection{Variable proper mass concept}
 
The following definitions and denotations are used. In Minkowski space of metric $\eta_{\mu \nu}$, any 4-vector $x^\mu$\ ($\mu=0, 1, 2 ,3$) is characterized by a time (temporal) component $x^0$ and a space (spatial) part, that is the 3-vector $x^i$ ($i=1, 2, 3$) in 3-space. The inner scalar product is defined ${\bf x\cdot x}=\eta_{\mu \nu}x^\mu x^\nu=x^\mu x_\mu=(x^0)^2-\sum_i (x^i)^2$. The position vector in the 4-coordinate Minkowski space is $x^\mu =(c_0t, x^i)$; the vector traces the trajectory of motion (the world line), which is not a straight line if a field is present. The corresponding proper velocity vector, which is tangent unit one, is a generalization of 3-velocity $v^i$: $u^\mu=dx^\mu/ds$, where $ds=\sqrt{ds^\mu ds_\mu}=c_0 d\tau$ is the arc length interval of world line $s$, $c_0$ is the speed of light in the absence of field, and the so-called proper time interval $d\tau= dt/\gamma$ is related to the coordinate time $t$ in the formula $v^i=dx^i/dt$ ( $d\tau$ is Lorentz invariant only in the absense of sources of field).

The 4-momentum vector is introduced as a generalization of the 3-momentum: $P^\mu=mu^\mu=m(dx^\mu/ds)$ with the obvious connection to $v^i$: $P^\mu=(m\gamma, \  m\gamma v^i/c_0)$. From this, the 4-momentum magnitude equals the proper mass $\sqrt{P^\mu P_\mu}=m$.

An important stage in Relativistic Mechanics is the introduction of Minkowski force $K^\mu$ (so far not specified) acting on a test particle of the proper mass $m$. In GR and conventional Relativistic Mechanics the proper mass is assumed to be constant $m=m_0$, so the dynamics equation has the form 
\begin{equation} 
dP^{\mu}/ds=d(m_0 u^\mu) /ds=m_0 du^\mu /ds= K^\mu 
\label{l}
\end{equation}
We change the above assumption and consider the proper mass being field-dependent $m=m(s)$ to allow for a non-zero tangent Minkowski force component $u^{\mu}(dm/ds)$
\begin{equation} 
K^\mu=dP^\mu/ds=u^{\mu}(dm/ds)+m(du^{\mu}/ds) 
\label{2}
\end{equation}

The question arises: how does one know whether the proper mass is constant (as mostly assumed in current field theories) or field dependent (as suggested in this work)? Our viewpoint is that the proper mass constancy assumption is the issue of theory physical foundations and subject to experimental falsification. It should be noted that the proper mass variability is not a new idea: it was discussed in classical books on relativity theory by Synge \cite {Synge} and Moller \cite {Moller} and occasionally later on in connections with field theories but did not draw much attention among physics community. We are going to confirm that the introduction of the field-dependent proper mass in the relativistic Lagrangean framework leads to a consistent relativistic mechanics.

\subsection{A relativistic generalization of static gravitational force}
\label{2.1}

Consider a test particle characterized by a field-dependent proper mass $m$. Let the particle be slowly moved at a constant speed along the radial direction in the $1/r$ static gravitational potential field due to a spherical source of a radius $R$ and a mass $M_0>> m_0$, where $m_0$ is a particle proper mass at infinity. Such an imaginary experiment can be done by means of an ideal
transporting device provided with a recuperating battery. Work on the particle of a variable proper mass is given by:
\begin{equation} 
F(r)dr=m(r)c_0^2 d(r_g/r), \  \  \ (r\ge R)
\label{3}
\end{equation}
where $r_g=GM_0/c_0^2$ is a gravitational interaction radius.
Since the gravitational force is compensated by a reaction from the transporting device, the particle must exchange energy with the battery in a process of
mass-energy transformation. So the change of potential energy is related to the
proper mass change:
\begin{equation}
dm(r)=-m(r)d(r_g/r), \ \ r\ge R
\label{4}
\end{equation}
and the proper mass of the particle is a function of $r$:
\begin{equation}
m(r)=m_0\exp(-r_g/r), \ \ r\ge R
\label{5}
\end{equation}
In a weak field approximation $r\ge R>>r_g$, we have a Newtonian limit, and still can retain the proper mass variation:
\begin{equation}
m(r)\cong m_0(1-r_g/r),
\label{6}
\end{equation}
As is seen from (\ref{5}), the proper mass tends to exhaust as $(r_g/r)$ rises,       
while a gravitational potential energy takes the form:
\begin{equation}
W(r)=-m_0c^2[1-\exp (-r_g/r)]
\label{7}
\end{equation}
and the force work is given by
\begin{equation}
F(r)dr=d W(r)=m_0c_0^2{\cdot} \exp (-r_g/r)d(r_g/r)
\label{8}
\end{equation}
The potential energy changes within the range $-m_0c^2\le W(r)\le 0$. Therefore, it
is limited by the factor $c^2$, and a divergence of gravitational energy is naturally eliminated. The same will be shown true for a particle in free fall. 

It is interesting to note that in the time of GR development, Finnish physicist G. Nordstroem \cite{Nordstroem} tried to develop an alternative gravitational mechanics and field theory. Obviously, he was aware of option (\ref{2}), in which the proper mass depends on a gravitational potential $\phi(r)$. In 1912-13 he considered a formulae $m(r)=m_0\exp (-g\phi)$ with some ``adjusting factor'' $g$. Having troubles with gravitational properties of light and inertial mass, he did not come to a consistent theory and abandoned work after Einstein's GR was published in 1915.  

From  (\ref{5}) it follows  that a predicted deviation from $1/r$ potential is noticeable near a source of high mss density, and it is not realistic yet to observe the effect in laboratories. Nevertheless, challenging experiments are in progress. In one of them, an alleged test of a supersymmetry theory prediction of the $1/r^2$ law violation is attempted with the use of a symmetric torsion pendulum  \cite{Hoyle}. The authors look for a quite large correction $[1+\alpha \exp{(-r/\lambda)]}$ in a direction, which is opposite to what we predict. Their assessment of the effect was obtained by conventional mechanics methods based on the gravitational force concept, in fact, similar to that of mechanical force: the kinetic energy gain $(\gamma-1)m_0$ is taken from an ``inexhaustible''  source. For this reason, potential energy is subject to $1/r$ divergence. We are motivated by the prediction of a new phenomenon, the proper mass exhaustion (\ref{5}) under strong field conditions. The phenomenon leads to a natural elimination of the divergence.

\subsection{Relativistic Lagrangean formulation of the problem}

For a particle of variable proper mass $m(s)$, $s=s({x_\mu})$, in a gravitational $1/r$ potential field, it is convenient to introduce {\it a proper Lagrangian} $L(s)$ in order to exploit the Minkowski force concept $K^\mu = -\partial{W}/\partial{x_\mu}$. (In fact, we cannot formulate the Lagrangian in terms of coordinate time $t$ since a relationship of Minkowski and ``ordinary'' forces is not known prior to proper Lagrangian study).   
A relativistic analog to the difference of kinetic and potential energy in our case is $(m_0-m)$ with $m u^\mu u_\mu=P^\mu u_\mu $. Because of the identity $u^\mu u_\mu=1$ and the source being stationary, the Lagrangian should not be an explicit function of $u^\mu$ or $s$. The proper mass must monotonously decrease as the particle approaches the source since a mass defect is associated with a growing binding energy. Yet, the potential field concept requires that $m\to m_0$, $W\to 0$ at infinity. In terms of the Noether's theorem, the $s$-translation symmetry, or the $t$-translation in $(t, x^i)$ coordinate system, is a manifestation of relativistic total energy conservation of a particle in a potential field.

We are going to study the time translation symmetry in Euler-Lagrange equations derived from the Hamilton's action principle. A relationship between any type of symmetry of a dynamical system with a conservation of corresponding quantity (Noether's current) is elegantly follows from the famous Noether's theorem. Her method works in a spirit of Hamilton's reformulation of Lagrangean mechanics. A trivial example is a classical system characterized by a set of generalized dynamical variables 
$[q(t),\ \dot q (t)]$ and a Lagrangian $L[(q(t),\ \dot q (t)]$, a system evolution is determined by the Euler-Lagrange equations  
\begin{equation}
\frac{d}{dt}[{\partial L}/{\partial \dot q}]={\partial L}/{\partial q}
\label{9}
\end{equation}
If the r.h.s. of (\ref{9}) is zero (the system has a $q$-symmetry), a quantity ${\partial L}/{\partial \dot q}$ is conserved. If the system additionally has the time symmetry, then 
$d L(q, \dot q)]/dt-[{\partial L }/{\partial q)}{\dot q }+ ({\partial L }/{\partial \dot q})({\partial  \dot q) }/{\partial t) })] = 0$. From this, in combination with (\ref{9}), the conserved Noether's current 
$ j=[\dot q ({\partial L }/{\partial \dot q}) - L]$ is derived. It characterizes a sum of kinetic and potential 
energy, the Hamiltonian $H=(T+W)$. 

Back to our problem: having the term $T=P^\mu u_\mu=m$ in the proper Lagrangian, one gets the Noether's conserved current $j=m_0-m+W=0$ (a change of kinetic energy equals a change of potential energy, their sum equals zero). It satisfies the requirement $(m_0-m)\to 0$, $W\to 0$ at infinity ($W\le 0$).
With the inclusion of rest mass, the conserved current is total energy, the Hamiltonian
\begin{equation}
H=m_0+T+W=m_0
\label{10}
\end{equation}
We shall return to this issue later in discussions of relativistic Euler-Lagrange equations.

\subsection{Equations of motion}

As was explained, the stationary Lagrangian is given by  
\begin{equation} 
L(s)=-m(s)-W(s)
\label{11}
\end{equation}
where $s=s(x^\mu)$ is a world line (arc)length, and a field is characterized by potential energy $W(s)$ (it is  negative for an attractive force). 
The Euler-Lagrange equations of motion follow from Hamilton's principle of the extremal action $S$
\begin{equation} 
\delta S=\delta \int_a^b L(s)ds=\int_a^b (\delta L) ds + \int_a^b  L d(\delta s)= \delta S_1 + \delta S_2 = 0 
\label{12}
\end{equation}
with a set of dynamical variables $x^\mu$ (the $s$ is not the one). Obviously, the proper mass $m(s)$ should not be considered an additional dynamical variable in a sense of the fifth degree of freedom. Thereafter, $m(s)$, 
$W(s)$, u(s), $s$, and $ds$ are subject to variation through independent variations of $x^\mu$. The proper velocity $u^\mu(s)$ as a function of dynamic variables $x^\mu$ will appear in the variational procedure, as well. 

It should be noted that the relativistic Lagrangean problem for a free particle motion was discussed in \cite{Landau},  \cite{Goldstein} with $W(s)=0$, the 
Lagrangian $L(s)=-m_0$ (in our denotations) and the action variation 
$\delta S=m_0 \delta \int_a^b ds=m_0\int_a^b d(\delta s)=0$. Clearly, this is a particular case of (\ref{12}).

From (\ref{12}) to continue, we have
\begin{equation} 
\delta S_1= \int_a^b (\delta L) ds=\int_a^b \frac{\partial L(s)}{\partial s}u^\mu\delta x_\mu ds                        
\label{13}
\end{equation}
\begin{equation} 
\delta S_2= \int_a^b L \delta(ds)=\int_a^b L \delta (u^\mu) dx_\mu=\int_a^b  L \frac{\partial u^\mu}{\partial s} \delta x_\mu ds                        
\label{14}
\end{equation}
\begin{equation} 
\delta S=\delta S_1+\delta S_2= \int_a^b \frac{d}{ds} (L u^\mu) \delta x_\mu ds=0                        
\label{15}
\end{equation}
Because variations $\delta x_\mu$  between the end points are independent for different $\mu$, the equality $\delta S=0$ in (\ref{15}) is possible if and only if 
\begin{equation}
\frac{d \left[L(s)u^\mu(s)\right]}{d s} =0
\label{16}
\end{equation}
With the Lagrangian (\ref{11}) substituted into (\ref{16}), we have Euler-Lagrange equations of motion
\begin{equation}
\frac{\partial \left[m(s)u^\mu(s)\right]}{\partial s}=-\frac{\partial \left[W(s)u^\mu(s)\right]}{\partial s}
\label{17}
\end{equation}
Having the additional equation of time-likeness of particle motion
\begin{equation}
 u^\mu u_\mu=1, \  \ u^\mu (du_\mu/ds)=0
\label{18}
\end{equation}
one is able to determine five correlated quantities $x^\mu(s)$, $m(s)$. 
Finally, one needs to introduce Minkowski force $K^\mu=-u^\mu \left ( {\partial W}/ {\partial s}\right)$ to get the desired equation of motion in terms of 4-momentum rate and Minkowski force
\begin{equation}
\frac{d}{ds}\left(m u^\mu \right) = K^\mu
\label{19}
\end{equation}
what is often intuitively written in conventional relativistic Physics. In the variable proper mass approach, however, there are actually two orthogonal (vector) equations in (\ref{19})  
\begin{eqnarray} 
u^\mu (dm/ds)=K^\mu_{tan}, & \  m (du^\mu/ds)=K^\mu_{per}
\label{20}
\end{eqnarray}
where  $ u^\mu (dm/ds)=K^\mu_{tan}= -u^\mu {\partial W}/{\partial s}$ is a tangential component, and  
$m (du^\mu/ds)= K_{per}^\mu =-W (du^\mu/ds)$ is due to a Minkowski force component acting perpendicularly to the world line. The two equations are coupled in a feedback manner through a varying proper mass. From the scalar product $P^\mu u_\mu$ and (\ref{18}), the following useful formulae are obtained: 
\begin{eqnarray} 
 K^\mu u_\mu = dm/ds, &  K^0 u_0 \ = dm/ds + \ K^i u_i & (i=1,\ 2,\ 3) 
\label{21}
\end{eqnarray}
which express an energy balance (a current in 4-space). The existence of two orthogonal solutions is a consequence of proper mass variability under force action. This is a new result, significance of which is seen in applications.

\section{The ${\bf 1/r}$ Gravitational Potential}

\subsection{Equations of motion}

For the practical use of results obtained in previous sections, one should express (\ref{19}) in terms of time-dependent 3-space coordinates $x^i(t)$ using a connection of proper/improper quantities $ds=c_0 dt/\gamma$ and the definition of $P^\mu$. The $t$ is a ``wristwatch'' time measured by an observer at rest with respect to the source but far away from it (ideally, at infinity), as discussed later.

The spatial part of (\ref{19}) is given by 
\begin{equation}
\frac{d}{dt}(\gamma mv^i)=F^i 
\label{22}
\end{equation}
with the relationship between Minkowski and ordinary forces acting on a test particle in 3-space
\begin{equation}
F^i={\frac{c_0^2}{\gamma}}K^i
\label{23}
\end{equation}
The second independent equation follows from the temporal part of (\ref{19}):
\begin{equation}
\frac{d}{dt}(\gamma m)=\frac{c_0}{\gamma}K^0
\label{24}
\end{equation}
which expresses the total energy rate of the particle in the field.  By definition of a conservative field, 
$K_0$, being a total energy rate, must be zero, hence, $\gamma m=C$. For the particle starting free fall from rest at infinity, $C=m_0$, $\gamma m=m_0$. This result will be later substantiated by considering the Noether's conservative current, which is recognized in (\ref{21}) or, equivalently
\begin{equation}
\frac{d}{dt}(\gamma m c_0^2)=F^i v_i+\frac{c_0^2}{\gamma}\frac{dm}{dt}
\label{25}
\end{equation}

Further we are to restrain ourself to the problem of free radial fall; an orbital motion is subject to a separate work. Thus, $dr(t)=c_0\beta(t) dt$, and (\ref{25}) becomes
\begin{equation}
\gamma d(\gamma m c_0^2)=\gamma F(r)dr+c_0^2 dm
\label{26}
\end{equation}
which is the total energy balance in a differential form. In fact, this is the Noether's conservative current discussed earlier in terms of proper quantities and now expressed in the  ($r,\ t$) coordinates in the differential form. It manifests a total energy conservation law for a particle in a spherical symmetric potential field: ``the conserved total energy'' equals a sum of ``the potential energy change due to gravitational force work'' and the corresponding  ``kinetic energy change'', where the total energy is $\gamma m=m_0$ in the considered case of free fall from rest at infinity. Therefore, the l.h.s. of (\ref{26}) is zero. 

Next step is to substitute the gravitational force expression (\ref{3}) into (\ref{22}) (or equivalently (\ref{26})) to find the proper mass function $m(r)$ taking into account the conservation $\gamma m=m_0$. Further we use a denotation for the ratio of values of the poroper mass at infinity and at point $r$: $\gamma_r=m_0/m(r)$ (the ratio plays an important role in our theory). With this, the equation for radial motion takes the form $m_0c_0^2 \gamma\beta d\beta=m_0c_0^2 d(r_g/r)$ with the dynamical solution to it
\begin{equation}
1/\gamma_r=m(r)/m_0=1-r_g/r, \  \ m(r)=m_0(1-r_g/r)
\label{27}
\end{equation}
where $r=r(t)$,  $\gamma_r (t)=\gamma[r(t)]$ that is, $\gamma m=m_0$ with $m[r(t)]$ as a function of $r$ in 
(\ref{27}).
It looks like a linear approximation (\ref{6}) of the static relation (\ref{5}) and consequently has a range restriction
$(r\ge R> r_g)$, discussed later. 
From this solution, kinetic energy as a difference of total and proper energy is 
\begin{eqnarray}
E_{kin}=m_0 c_0^2-m(r) c_0^2 =m c_0^2 (r_g/r) & \ (r\ge R> r_g)
\label{28}
\end{eqnarray}
while the sum of kinetic and potential energy changes equals zero what makes the total energy $E_{tot}=m_0 c_0^2$. By finding the specific function $m(r)$ (\ref{27}) from equations of motion, we confirmed the Noether's current concept (\ref{10}) and the constancy $\gamma m=m_0$.

If the particle in radial fall has kinetic energy at infinity $E_0=\gamma_0 m_0^2$ then, due to the total energy conservation, $m_0$ should be replaced by $\gamma_0 m_0$; correspondingly, the equality $\gamma(t)=\gamma_r(t)$ should be replaced by $\gamma (t)=\gamma_0 \gamma_r (t)$, where $\gamma_r=m_0/m(r)$, $r=r(t)$ as before, $\gamma_0=(1-v_0^2/c_0^2)^{-1/2}$, $v_0$ is the radial speed at infinity. Then (\ref{28}) becomes 
\begin{eqnarray}
E_{kin}= m c_0^2(\gamma_0-1)+r_g/r
\label{29}
\end{eqnarray}
However, the results of this subsection is not final. The matter is that we need to take into account the mass defect in the source. This effect was so far ignored, and we are going to correct our results in the next subsection.

\subsection{Correction for the source mass defect, and final results.}
\label{3.2}

The requirement of $(r_g<R)$ in (\ref{28}) precludes the proper mass from reaching a zero value in the exterior region when $m\to 0$ at $r\to r_g$. The problem is caused by the simplified concept of the gravitational radius $r_g=GM/ c_0^2$: we did not bother ourselves with finding both exterior and interior solutions and matching both at $r=R$. The simplification concerns a binding energy of the sphere (a mass defect): the latter cannot be found unless the interior solution for $r<r_g$ is known. In other words, one need to take into account the fact that $M\ne  M_0=\sum_i {m_{0i}}$, where $m_{i0}$ are proper masses ``at infinity'' of particles comprising the sphere. The difference is a self-binding energy $\Delta M=M_0-M$. 

In fact, the interior solution needs material structure taken into account, so a solution of the prolem will be approximate anyway. Thereafter, we are going to reformulate the problem in terms of $r_{g0}=GM_0 / c_0^2$ with the correction for the mass defect. An approximate way to do it would be to introduce a spacial factor
$M_0/M=m_0/m=\gamma_r (r)$. Then, the gravitational force takes the form  
\begin{eqnarray}
F(r)dr=GM_0 (m^2/m_0)dr(1/r)=m_0c_0^2(m/m_0)^2 d(r_{g0}/r) 
\label{30}
\end{eqnarray}
The correction ensures physical requirement $(m(r)>0)$ in the whole range $(r>R)$ and a boundary junction of exterior solution $m(r)$ at $(r\ge R)$ with that at the surface $r=R$ without actual finding the interior solution. Further on, we drop the lower zero index in $r_{g0}$ and use the previous denotation $r_g=GM_0/c_0^2$ for the gravitational radius having a new meaning. The introduction of the additional factor $\gamma_r=m_0/m$ in the source term is an approximate way to account for the source self-binding effect in order to correct a radial dependence of an exterior field under strong field conditions. 

All things considered, the equation (\ref{22}) takes the form 
\begin{eqnarray}
\gamma^2 \beta d\beta=d(r_{g}/r) 
\label{31}
\end{eqnarray}
and the dynamical solution is: 
\begin{eqnarray}
1/\gamma_r= m/m_0=\exp{(-r_{g}/r)}, \ &  (r\ge R) 
\label{32}
\end{eqnarray}
It coincides with the static solution  (\ref{5}) (what is not necessarily expected). Having kinetic energy term $\gamma_0$ been accounted for from the condition at infinity,
we have a final set of formulae:
\begin{eqnarray}
\gamma=\gamma_0\gamma_r=\gamma_0\exp(r_g/r), & 
\beta(r)=\left[1-(1/\gamma_0^2)\exp(-2r_g/r)\right]^{1/2} 
\label{33}
\end{eqnarray}
and squared norms of the 4-momentum $P^\mu=m(\gamma,\gamma\beta,0,0)$  and the 4-coordinate vector 
$\Delta x^\mu=c_0\Delta\tau(\gamma,\gamma\beta,0,0)$
\begin{eqnarray}
(c_0 m)^2=(c_0 \gamma m)^2- p^2  
\label{34}
\end{eqnarray}
\begin{eqnarray}
 (\Delta s)^2=(c_0 \gamma \Delta \tau)^2- (\Delta r)^2  
\label{35}
\end{eqnarray}
Relations will be used further: $\gamma=\gamma_r \gamma_0$, $\gamma m=\gamma_0 m_0$, 
$\gamma \Delta\tau=\gamma_0 \Delta t_0$, $p=c_0 \gamma \beta m=c_0 \gamma_0 \beta m_0$, 
$\Delta s=c_0 \Delta\tau$,  
$\Delta r= c_0 \gamma\beta\Delta\tau= c_0 \gamma_0 \beta\Delta t_0$ ($t_0$ is the ``coordinate'' time measured by the rest observer at infinity; it is usually denoted $t$, as discussed later). 
 
Formulae for total, kinetic, and potential energy are:
\begin{equation}
(E_{tot}/c_0)^2=(\gamma_0 m_0 c_0)^2 =p^2+ (m c_0)^2 
\label{36}
\end{equation}
\begin{equation}
E_{kin}(r)=E_{tot}-m c_0^2=m_0 c_0^2 \left[\gamma_0-\exp(-r_g/r)\right] 
\label{37}
\end{equation}
\begin{eqnarray}
W(r)=-m_0 c_0^2\left[ 1-\exp{(-r_g/r)} \right] \ , &   (r\ge R)
\label{38} 
\end{eqnarray}
It is seen that \  $-m_0c_0^2 \le W \le 0$. When $m_0<<M_0$, the kinetic energy emerges solely due to the change of the proper mass of a test particle in a field, and the proper mass ``exhaustion'' under strong field conditions takes place. We want to emphasize again that
the divergence is eliminated for an arbitrary mass density of the source and a however strong field.

Under weak-field conditions $r_g/r<<1$, we have
\begin{equation}
\gamma=\gamma_0(1+r_g/r), \ \ \beta=\left[1-(1-r_g/r)/\gamma_0) \right]^{1/2} 
\label{39}
\end{equation}
\begin{equation}
E_{kin}=m_0 c_0^2(\gamma_0-1+r_g/r) 
\label{40}
\end{equation}
\begin{equation}
W(r)=-m_0 c_0^2 (r_g/r),  \  \phi(r)=W(r)/m_0 c_0^2=-(r_g/r)  
\label{41}
\end{equation}
and the Newtonian limit $E_{kin}=mv^2 /2$.

It should be noted that equations of test particle motion can be presented in the form independent on test particle mass (it reflects gravitational/inertial mass equality). The argument can be a coordinate time 
$t$, or a position 3-vector $x^i=x^i (t)$. Obviously, same facts take place in conventional mechanics. However, we have an additional equation (\ref{32}) to determine the proper mass function.

Clearly, our results and conventional ones differ at high energies due to the difference in concepts of relativistic mass and, correspondingly, potential energy. A particle to be accelerated by a force at distance needs to be bound. The binding energy in our philosophy is a real mass defect $(m-m_0)$ limited by the proper mass value. It makes the force weaken as $r\to r_g$ so that no infinities arise. In the concept of proper mass constancy, the concept of binding energy is not clear; it looks like the particle gets bound while acquiring kinetic energy from unlimited field energy, so both the binding and kinetic energy, in principle, are unlimited.

\section{Lagrangian symmetry, Noether's theorem, and energy conservation}

\subsection{Time-translation symmetry, and energy conservation}
\label{4.1}

In order to study the Noether's current in more details, let us go back to (\ref{34}), (\ref{35}) to consider a world line in the 4-momentum space in a manner as we do in the 4-coordinate space, and compare 4-vector norms:
$ \Delta S(r)=|\Delta {\bf x|}$ and $ \Delta S_p (r)=|\Delta {\bf P|}$ of the coordinate vector 
$\Delta x^\mu=c_0 \Delta t (1,\ \beta,\ 0,\ 0)$ and the momentum one $c_0 P^\mu= c_0 m_0 (1,\ \beta,\ 0,\ 0)$, respectively. Thus, we need to compare how proper mass and time behave. In the case of a radial motion from rest at infinity, the Lorentzian norms are:   
\begin{equation} 
\Delta S(r)=\left[(c_0 \Delta t_0)^2- (\Delta r)^2\right]^{1/2}=c_0\Delta t_0/\gamma_r=c_0\Delta\tau(r) 
\label{42}
\end{equation}
\begin{equation}
\Delta S_p (r)=\left[(c_0 m_0)^2- (p(r))^2\right]^{1/2}=c_0 m_0/\gamma_r=c_0 m(r) 
\label{43}
\end{equation}
where  $\Delta S(r)=\Delta s(r)=\gamma(r) \Delta\tau(r)$ is the world line interval in (\ref{35}), $r=r(t)$,  $c_0 \beta(r)=\Delta r/\Delta t=\Delta r'/\Delta\tau(r)$. The operational meaning is, as next. $\Delta r=c_0\beta \Delta t_0 $ is measured by a ``far-away observer'' at rest with respect to the source. She determines $\beta$ from measured $\Delta r$ per a constant time interval $\Delta t_0$ by the time-of-flight technique with the use of standard clocks and rods. Thus, we term $t=t_0$ with zero subscript ``a far-away time'', also called ``a coordinate time''. Next quantities are the contracted radial interval  $\Delta r'=\Delta r/\gamma$, and the world line interval $\Delta s(r)=c_0\Delta\tau(r)$ both measured by a comoving observer. The contraction is a pure SR kinematical effect. It is seen that  $\Delta s(r)=c_0 \Delta \tau(r)$ is not invariant. Notice that $\beta=\Delta r/\Delta t_0=\Delta r'/\Delta \tau$.

From measurements of the speed $\beta$, the gravitational time dilation effect can be obtained. The latter is associated with the proper mass dependence on the gravitational potential $m(r)=m_0\exp(-r_g/r)$. The corresponding frequency of atomic clock of the proper mass $m(r)$ is proportional to the proper mass: $m(r)c_0^2=h f(r)$, where $f(r)=1/T(r)$ is a relationship of $f(r)$  with the proper period 
$T(r)=\gamma_r \Delta t_0$; this is a rest time interval at point $r$ (that is, in the observer's reference frame). Hence, $T(r)=\Delta t_0$ at infinity. The clock at a deeper potential level $r_2\to r_1$, $r_2>r_1$ will slow down by the factor $\gamma_r$ in agreement with observations. Therefore, one needs the factor $\gamma_r=(1-\beta^2)^{1/2}=\exp{(r_g/r)}$ from measured values of $\beta$ for $\gamma_0=1$ to find $m(r)=m_0/\gamma_r$, $f(r)=f_0/\gamma_r$. There is a useful relationship $\Delta\tau (r) T(r)=\Delta t_0^2 $.  It becomes clear that the assumption of the proper mass constancy in the SR-based mechanics would result in a failure of a gravitational time dilation prediction. This is one of the reasons to discard the assumption of proper mass constancy in the SR-based gravitational theory. Changing the assumption makes a desired difference.

One can recognize a new conservation symmetry by examining 4-vector components in (\ref{42}), (\ref{43}) rearranged in (\ref{44}); $\gamma_0$ is put equal to unit for simplicity there. Given conditions at infinity, conserved quantities are seen on the l.h.s. of each equation in (\ref{44}):
\begin{equation}
[c_0\Delta t_0]^2=[\Delta S (r)]^2 + [\Delta r]^2,\   \  
(c_0 m_0)^2 = [\Delta S_p (r)]^2 + [p(r)]^2  
\label{44}
\end{equation}
Instead of hyperbolic rotation in  metric (+, -, -, -), a real rotation symmetry emerged in a quasi-Euclidean geometry of signature (+, +, +, +). The constant radius of rotation is 
$c_0 \Delta t_0$ and $c_0 m_0$ in coordinate and momentum space, correspondingly. The rotation angle $\theta$ is determined by $\sin\theta=\beta [r(t)] $,\ or identically \ $\cos\theta=1/\gamma=\gamma [r(t)]$. Compare it with an imaginary angle $\psi$ of hyperbolic rotation: $\cosh\psi=\gamma$, $\sinh\psi=\gamma\beta$, 
$\tan\psi=\beta$. Hence, $\sin\theta=\tanh\psi=\beta$.

\subsection{Total energy conservation law and dynamical complementarity principle}

The new (real rotation) symmetry ensures the total energy conservation law in the approach of the variable proper mass concept. It can be shown that a similar symmetry takes place under general conditions at infinity when $(\gamma_0>0)$ or  $(\gamma_0<0)$; the case of a negative initial kinetic energy at infinity means that the test particle is dropped at some finite point $r>R$ where the potential is not zero. The usual interpretation of the energy conservation is known in terms of Nether's conserved mass-energy current in the momentum space. Our finding is that there is a similar conserved current in the coordinate space. It corresponds to the constant time rate recorded by a far-away atomic clock. Therefore, there are two, equivalent, symmetries in $P^\mu$ and $x^\mu$ spaces under gravitational dynamics conditions.  This fact is consistent with SR Kinematics. Recall that the Klein-Gordon equation is derived in the SR Kinematics framework with the relativistic de Broglie wave concept introduced. The latter includes such quantities as the 4-phase $\phi=(\omega t -{\bf k\cdot r})$, where the 4-wave vector $k^\mu$ is proportional to $P^\mu$: $(\hbar/c_0) k^\mu = \hbar/c_0(\omega,\ {\bf k}) = (E/c_0,\ {\bf p})$, where $E=m c_0^2=\hbar \omega=hf$. The following, proportional to the phase, scalar product  is Lorentz invariant:
\begin{eqnarray}
c_0P^\mu \Delta x_\mu=h \  \ \mbox{or} \  \ c_0 P^\mu  x_\mu=N h 
\label{45}
\end{eqnarray}
where $N$ is a number of wavelength (clock ticks).

It is not surprising that our generalization of the de Broglie concept to the {\it relativistic gravitational dynamical problem} came from the Lagrangean problem formulaton in the form of the quasi-Euclidean representation (\ref{44}). As a relult, the invariance of the 4-phase, or the scalar product analogous to (\ref{45}), takes place in the quasi-Euclidean metric. The operational meaning of it is clear and it is the same in SR Kinematics and Dynamics: all observers agree to use standard atomic clocks of the proper mass $m_0$ at infinity (see subsection \ref{4.1}). The clock is considered a quantum oscillator in the de Broglie wave concept; consequently, $\Delta t_0$ and $f_0\sim m_0$ are reciprocal quantities.  

The fact of invariance (\ref{44}), (\ref{45}) in the quasi-Euclidean dynamical metric is called further  "the dynamical complementarity principle" due to its significance in our study. The  quantum de Broglie concept is seen to be naturally embedded in our SR-based gravitational dynamics before a field theory development. Some other issues relevant to the problem are discussed in \cite{Vankov2}. We believe that the real rotation symmetry is a true  law of Nature to be confirmed by observations. The law reflects the idea of mass and time unity and enables us to gain into a new insight of physical and philosophical concepts of matter and time. 

\subsection{Graphical illustrations, and lessons}

A brief comment is needed before discussing graphical illustrations of the free fall problem. In SR textbooks, the Lorentz kinematical transformation is usually illustrated by a straightforward picture of hyperbolic rotation. This would be a trigonometrical rotation in a complex plane by an imaginary angle $\psi$, $\tan {\psi}=\imath {\beta}$; optionally, it can be shown as a hyperbolic rotation in a real plane so that $\tanh {\phi}=\beta$, hence, $\tan {\psi}=\tanh {\phi}$. The idea in both variants is to show in the graph the invariant Lorentz norm $\Delta s$ {\it as a rotating radius}. 

Our graphic presentation is different and has more physical sense for us. There are three terms depicted in each graph in a real plane: ``spatial part'' versus ``Lorentzian norm''. The picture presents the Lorentzian quadratic metric: ``squared Lorentzian norm'' = ``squared temporal part'' - ``squared spatial part'', and at the same time the quasi-Euclidean one: ``squared temporal part'' =  ``squared Lorentzian norm''+ ``squared spatial part''. In the second case, a ``temporal part'' (not the Lorentzian norm) rotates in a real plane by a real angle $\theta=tan^{-1} (\gamma\beta)$. It is possible now to illustrate the norm invariance in usual Lorentz-boost transformations (the case of inertial motion) as well as a real rotation in the case of free radial fall. In Fig.\ref{Sym} each graph is presented equivalently in $(p,\ m)$ and $(r,\ \tau)$ planes of $P^\mu$ and $x^\mu$ Minkowski spaces, correspondingly. 

The case of the pure (no field) Minkowski space is considered firstly. There are three graphs $a)$, $b)$, and $c)$, which are different in a type of constraints imposed on Lorentz {\it kinematical} transformations in the 4-coordinate and 4-momentum spaces.



\begin{figure}[t]
\includegraphics{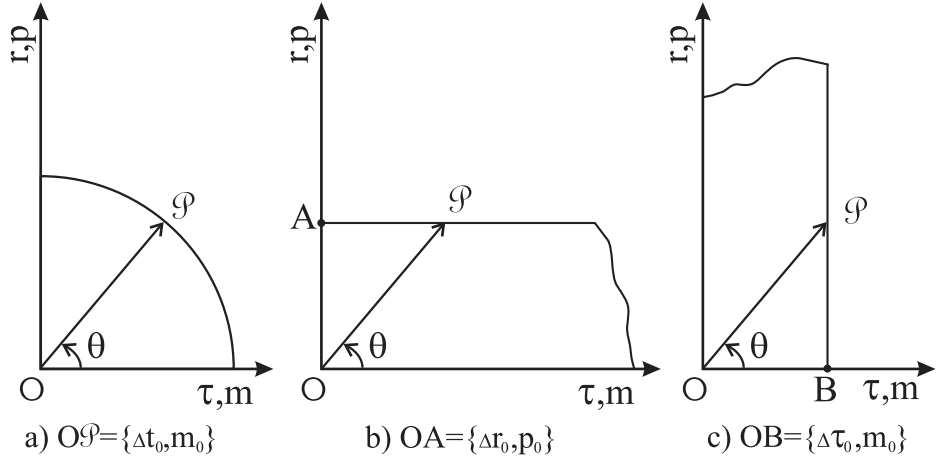}
\label{Sym}
\caption{\label{Sym}  4-vector kinematical and dynamical  rotation. $\theta=\tan^{-1}{(\gamma\beta)}$. 
{\it Graph a) SR Dynamics ($1/r$ potential field): real rotation symmetry. \  \  \  \  \ 
Graph b) Spatial part is fixed. There is no symmetries or invariance.\  \ \  \  \  \  \
Graph c) SR Kinematics (no field): proper mass and time Lorentz invariance.}}

\end{figure}

{\it Graph $c)$} presents a family of world lines with the parameter $\beta$ in pure (no field) Minkowski space. This is the case when observers travel with different speed provided the travel proper time $\Delta\tau_0$ measured by the traveling observer is fixed. ``The staying at home'' and traveling observers agreed to use standard atomic clocks to verify $\Delta\tau$ constancy. Consequently, both the proper time and the proper mass in the family of world lines are Lorentz invariant. Lorentz invariance does not takes place in other than {\it case c)} situations, as seen next.
   
{\it Graphs $b)$} describes the problem of travel with a speed $\beta$ (as a parameter) from $O$ to $A$ of a fixed distance (this is a constraint), $OA=\Delta r=\Delta r_0=c_0 \beta\Delta t$ in a pure Minkowski space. 
{\it Graph $a)$} is the case of the constraint $\Delta t=\Delta t_0$ (the travel time $\Delta t=\Delta t_0$ is fixed in the far-away observer's coordinate system). In both cases, it is not possible to make an agreement to use standard clocks: both the proper time and the proper mass depend on $\beta$ (they are not Lorentz invariant, and the kinematical complementarity does not hold).

Now, let us discuss our {\it SR-based dynamical problem} of free radial fall in the {\it graph $a)$}, $m(r)=m_0/\gamma_r$, $r=r(t)$. 
The graph illustrates the quasi-Euclidean representation of Minkowski 4-vectors. The family is produced in a single experiment with one freely falling particle, the world line of which is divided into small adjacent intervals (partitions), the Lorentzian norms $\Delta s =c_0 \Delta\tau $ provided $\Delta t=\Delta t_0$. Conditions at infinity are fixed: $\theta=0$, $\Delta\tau=\Delta t=\Delta t_0$; ($\gamma_0=1$ for simplicity). The dynamical complementarity principle and the time translation symmetry hold. The Lorentzian norm and the proper mass are functions of 
$\beta$, while the vector $OP=c_0\Delta t_0$ is a conserved temporal component. The graph shows a real rotation of $OP$ with a constant radius of rotation (the Noether's current) 
$OP=c_0 \gamma\Delta\tau=c_0 \Delta t_0$ in $x^\mu$ space and 
$OP=\gamma m =m_0$ in $P^\mu$ space. The angle $\theta(\beta)$ is a function of dynamic variables $x^\mu$. It characterizes an instantaneous state of a freely falling particle at an instant $t$: $\theta (r)=\sin^{-1} \beta (r)$, $r=r(t)$. Obviously, {\it graphs $b)$} and {\it $c)$} are not relevant to our problem, though $a)$ reduces to $c)$ at infinity. 

\subsection{Lessons and some discussions}
\label{4.4}

Let us discuss some lessons drawn from graphs and make a comparison of predictions within SR-based gravitational dynamics of point particle having constant-versus-variable proper mass.

1. {\em Kinematics and dynamics invariants}.
There are two categories of constant relativistic physical quantities. The first category relates to the hyperbolic rotation in pure Minkowski space. The constancy is due to constraints imposed on the Lorentzian vector representation. Only the constraint $c)$ leads to the kinematical complementarity principle and  Lorentz  invariance under $\beta$-boost transformations in 4-coordinate and 4-momentum complementary spaces. The Lorentz invariance is generated by the translation symmetry of a 4-point in Minkowski $x^\mu$ and $P^\mu$ space at the same time. An attempt to construct an ``extended'' Lorentz group without respecting the complementarity principle would mean the abuse of the Minkowski space concept.


One should distinguish the first (kinematical) category of Lorentz invariant quantities from the second (dynamical) category of conserved (unchanged in time) quantities in a Lagrangian system, the Noether's theorem deal with (as in {\it case $a$}). Our object under investigation is a 4-vector undergoing an evolution in the Lagrangian dynamical system in the Minkowski space. One can think of $\Delta s$ as an ``instantaneous image'' of the proper 4-position vector $OP$ tracing a small linear interval $s\to (s+\Delta s)$ on the curved  world-line $s$ in the 4-coordinate space or $m\to (m+ \Delta m)$ in the 4-momentum space. In the picture $a)$, the interval is a $P$-projection on the $\tau$-axis, (or on the $m$-axis) and it is not constant: it gets smaller as the particle approaches the source. However, the interval $\Delta t=OP$ is preserved. To find the proportion $\Delta t_0/\Delta\tau(r)= T(r)/\Delta t_0$ and the time interval $T(r)$ referred to the gravitational time dilation, one needs to draw the tangent line at the point $P$ to the intersection with the horizontal axis.

2. {\em Noether's conservative currents and dynamical symmetries}. 
The picture $a)$ illustrates SR-based gravitational dynamics, essential part of which is the field-dependent proper mass concept. The conservation takes the form of rotation symmetry in a real plane in a quasi-Euclidean 4-space, and it is associated with the Noether's conserved current due to the time translation symmetry, as in classical mechanics. 

Descriptions of free fall in space-time and in the 4-momentum space are formally identical. Indeed, the final equation of motion (\ref{31}) does not contain a mass of a test particle (as in classical mechanics). The parameter of physical importance is the gravitational interaction radius $r_g$ in the gauge factor $\gamma_r=r_g/r$: the factor determines Minkowski space deformation via space-time and mass-energy rescaling. On the one hand, the source  causes proper mass variation under Minkowski force action (the momentum space curvature). On the other hand, it makes the world line curved (the coordinate space curvature). Certainly, the two currents would follow from the Noether'e theorem, provided the complementarity principle was stated in the Lagrangean problem formulation.

3. {\em Two (complementary) currents}. 
``Two currents'' means that in our originally formulated relativistic Lagrangian, the proper mass $m$ can be replaced by the complementary  quantity $\Delta\tau$ to allow the Minkowski force coming to the scene in the momentum $K_m^\mu$ representation (the proper mass being affected by field), or coordinate $K_{\tau}^\mu$ representation (the proper time pace being affected by field) with the equivalent outcome. To agree on this proposition, one should think about acted by force particles in a broader concept of atomic clocks that is, influenced by field quantum oscillators probing both mass/energy and space/time local metric in comparison with the constant background at infinity. (In other words, a consideration of the de Broglie wave propagation in a gravitational field is suggested). Consequently, we deal with a complementary Lagrangean formulation of the problem resulted in two complementary solutions: 
\begin{eqnarray}
m=m_0\exp{(-r_g/r)}, & \Delta\tau=\Delta t_0\exp{(-r_g/r)} 
\label{46}
\end{eqnarray}
obtained in Sections 2 and 3 without emphasizing the fact of a double-fold formulation. To check if it is true, just put $m=m_0/\gamma$ and $u^\mu=(\gamma, \gamma\beta, 0, 0)$ for $K_m$ in (\ref{19}), and do the same with $\Delta\tau=\Delta\tau_0/\gamma$ for $K_{\tau}$. Finally, the same equations of motion are derived.

4.  {\em Alternative versus conventional theory comparison}.  In  subsection \ref{3.2} (see comparisons also in  \ref{2.1},  \ref{4.1} ), we concluded that the proper mass constancy in gravitational relativistic mechanics leads to the potential energy divergence. Now we are able to examine another issue:  the role of the proper mass concept in Lorentz invariance of 4-vector norms (\ref{42}) and (\ref{43}). Obviously, if the proper mass is assumed constant, then $\Delta S_p=c_0 m_0$. Moreover, the world line length (or the proper time interval) must be constant, $\Delta S=c_0 \Delta\tau_0$, especially if we persue the idea of complementarity 
$m_0 c_0^2=h/\Delta\tau_0$. Consequently, one faces the problem of gravitational time dilation prediction. As is known, the effect is described in the GR framework in agreement with observations. It should be noted that the proper velocity vector $u^\mu$ remains Lorentz invariant in Minkowski space regardless of the proper mass concept. This is not surprising because the vector characterizes the time-like character of massive particle motion in Minkowski space in the form of hyperbolic rotation identity $1=\gamma^2-(\gamma\beta)^2$. 

One needs to be sure about differnet terms  ``invariance'' (constancy of a vector norm, a scalar), and "covariance" (the 4-vector is Lorentz covariant if it transforms under a given representation of the Lorentz group).

Let us sum up the difference between conventional and our (alternative) SR Mechanics.

1)	Conventional theory. 

The 4-velocity $u^\mu=(\gamma, \gamma\beta^i)$, $sign(+, -, -, -)$
is constructed from the identity  $\gamma^2-(\gamma\beta)^2=1$; hence, Lorentz (local) invariance of the norm is a geometrically trivial invariance, reflecting hyperbolic 4-rotation for particle motion having the time-like character (unlike a photon). Physically, it is a fundamental property of Minkowski space. Notice, the 4-velocity changes a direction along world line in the presence of field because the world line is not a straight line. 

The 4-momentum is constructed by multiplying the 4-velocity vector by a scalar, the proper mass: $P^\mu=m u^\mu$. Hence, the norm is $m$, and in the conventional theory, it is $m=m_0=const$. The proper mass invariance is geometrically trivial: it reflects the proper mass constancy assumption plus identity due to 4-velocity being a unit (tangent) vector. On the other hand, the proper mass invariance is fundamental because the proper mass constancy assumption is a part of theory foundations.
The energy conservation takes the form: $\Delta PE + \Delta KE =0$, where $\Delta PE$ is a negative field energy. 

2) Alternative theory.

The initial Lagrangean problem is formulated in terms of Minkowski force related to the covariant (unit) 4-velocity $u^\mu (\gamma, \gamma\beta^i)$. 
The assumption is a revised proper mass concept: the proper mass is a function of field strength $m(r_g/r)$. Using the field conservativeness, we derive the Euler-Lagrange equations along with the function $m(r_g/r)/m_0=\exp(-r_g/r)=1/\gamma(r)$. The 4-velocity norm remains Lorentz invariant but the 4-momentum is trivially not Lorentz invariant (because of the assumption, which is the part of alternative theory foundations).

The Noether's conservative currents are found in the "quai-Euclidean space" ($M^*$-space for short), which is related to the Minkowski space through the 4-velocity. Now, the covariant in $M$ and $M^*$ unit 4-velocity is constructed in $M^*$ : ${u^*}^\mu (1/\gamma, \beta^i)$, $sign(+, +, +, +)$. The 4-momentum is 
${P^*}^\mu=m(r_g/r) {u^*}^\mu$, and the norm of $P^*$ is Lorentz invariant in $M^*$: the norm is $\gamma m=m_0$, the proper mass at infinity (constant).
The energy conservation takes the form: $\Delta PE + \Delta KE =0$, where $\Delta PE$ is a negative proper mass defect (negative binding energy).

The  situation is illustrated in graphs. If the proper mass is taken  constant (conventional theory), we lose the gravitational time dilation (found in GR) and get divergence. 
Therefore, a denial of proper mass Lorentz invariance in Minkowski space perturbed by sources is physical (compelling)  necessity in our philosophy rather than ``sacrifice''. However, terms should be clarified: we need to say ``vector norm invariance due to such and such symmetry''.

\medskip
In the next Section, we discuss the photon problem in the Minkowski (deformed) space. Instead of GR ``curved space-time field'', the more appropriate in SR Mechanics concept is introduced: ``gravitational refracting medium''.


\section{A photon in the gravitational field}

Unlike the particle, the photon does not have a proper mass; its total mass is solely a kinetic one. One has to look for conserved quantities in the photon metric taking into account the photon SR kinematics \cite{Vankov3}. Instead of detailed analysis, we simplify the problem by considering a photon emitter/detector at rest with respect to the source and making use of the fact that any photon in flight in a gravitational field is characterized by the two conserved quantities: an energy (frequency) and an angular momentum (the latter is out of consideration here). 

Thus, we assert that the energy (frequency) of the photon emitted at any point does not change during its travel in a gravitational field.  From the concept of the atomic clock, it follows that the frequency $f_{ph}$ at the instant of emission must be proportional to the frequency of an atomic clock-emitter $f(r)=m(r)c_0^2/h=f_0\exp(-r_g/r)$, that is, the {\it emission} frequency is field dependent. Therefore, the momentum (or the wavelength) and the speed of light will proportionally change with respect to those values measured by the far-away observer in experiments with the standard photon from her clock-emitter. All said above is sufficient for the determination of photon gravitational properties in the model, in which Minkowski space filled with field is considered a transparent refracting medium.

The next set of formulae describe characteristics of the photon detected at a point $r$, if emitted at a point $r'$.
\begin{equation}
f_{ph}(r'\to r)=f_0\exp(-r_g/r')
\label{47}
\end{equation}
where $f_0$ is the photon frequency at infinity; the photon does not change the initial (emission) frequency during its flight. The photon speed (the speed of light) is
\begin{equation}
c_{ph}(r'\to r)=c_0\exp(-r_g/r)
\label{48}
\end{equation}
So far, we consider results valid for all frequencies (there is no dispersion); hence, a photon and light propagate similarly. The speed of light at detection point $r$ does not depend on a point of emission $r'$. Consequently, the photon wavelength is
\begin{equation}
\lambda_{ph}(r'\to r)=\lambda_0 \exp(r_g/r'-r_g/r)  
\label{49}
\end{equation}
It follows that the photon wavelength at any point of emission equals the wavelength at infinity $\lambda_0$. Finally, the proper period of a resonance line of atomic clock is
\begin{equation}
T_{res}(r')=1/f_{res}(r')=T_0 \exp(r_g/r')
\label{50}
\end{equation}
All quantities with ``zero'' subscript are measured at infinity. The speed of light is influenced by the gravitational potential according to (\ref{48}); further a dimensionless form is used 
\begin{equation}
\beta_{ph}(r)=c(r)/c_0=\exp(-r_g/r)
\label{51}
\end{equation}
This is the speed of light wave propagation. Physical processes described by the above formulae are time reversal in accordance with the energy conservation. Thus, the gravitational time dilation and the red shift are due to the field dependence of the emission frequency and the speed of light, provided the photon energy being conserved.

It is seen that the speed of light is constant on the equipotential surface $r=r_0$, and it may be termed a tangential, or arc speed. One can define also the radial (``coordinate'') speed $\tilde \beta_{ph}(r)$
\begin{equation}
\tilde \beta_{ph}(r)=\beta_{ph}(r)(dr/d\lambda)=  \exp(-2r_g/r)
\label{52}
\end{equation}
Under weak-field conditions, it coincides with the corresponding GR formula. 

We conclude that the photon propagates in a gravitational field as in a refracting medium with the index of gravitational  refraction $n_g=1/\tilde \beta_{ph}$. The refraction concept was discussed in the GR literature (see, for example  \cite{Moller}, \cite{Fock}, \cite{Fischbach}). It should be noted that there is no evidence that a static electric or magnetic field alone would affect the speed of the photon. 


\section{Predictions and Observations}

GR tests are related to weak-field conditions and usually presented in literature as a solid GR gravitodynamics confirmation of the curved space-time concept \cite{Clifford, Weinberg}. In fact, under those conditions of ``near-Newton'' limit, a behavior of a photon and atomic clock in our approach is similar to that in GR (in spite of different space-time philosophy). How well our approach fits all observations is a special issue; many details need to be further investigated. Here we are able to make only a brief review of basic facts.

\medskip
{\it 1. The gravitational red-shift and time dilation}

The term ``red-shift'' means that the wavelength of a photon emitted by an
atomic clock at some point of lower potential appears to be increased when
detected at some point of higher potential. Our interpretation of the red-shift was explained earlier: the effect is due to a combination of the gravitational shift of the emission-detection resonance line and the dependence of the speed of light on field strength while a frequency of a photon in flight being constant and equal to the emission frequency (47-50). The latter is proportional to the field dependent proper mass $f\sim m$ what causes the gravitational time dilation. This interpretation is consistent with total energy and angular momentum conservation laws in the field-dependent proper mass concept. 


\medskip
{\it 2. The bending of light}

The bending of light is due
to the ``gravitational refraction''. We obtained the index of gravitational refraction consistent with that in GR, therefore, we predict the correct bending effect (calculations of the effect with the use of a refraction model are presented in textbooks). We also verified that the effect is consistent with the photon angular momentum conservation.  

\medskip
{\it 3. The time delay of light flight}

The time delay effect was measured in radar echo experiments with electromagnetic
pulses passing near the Sun. The effect can be calculated by 
integrating the time of light travel over the path with the field-dependent coordinate speed
(\ref{52}); the result will be equivalent to GR predictions. 

\medskip
{\it 4. Planetary perihelion precession and other astronomical observations}

This problem is related to a particle orbital motion in a gravitational field. It adds nothing new to our conclusion about absence of numerical difference in predictions of weak-field effects in the alternative versus conventional theory.  The perihelion precession can be assessed in GR by comparing radial and orbital frequencies in the Schwarzschild metric under weak-field approximation or in the post-Newtonian parameterization model. In the alternative approach, the corresponding physical treatment is equivalent to that in the effective potential model, in which dynamical quantities of orbital motion are influenced by the first-order field dependence of the proper mass in the Minkowski space.



\medskip
{\it 5. A particle in free fall in a gravitational field}

This is the case when we can compare predictions under high energy conditions. 
According to GR \cite{Misner}, a relative
speed of a particle in a radial fall is described by 
$\beta(r)=(1-2r_g/r)[1-(1-2r_g/r)/\gamma_0^2]^{1/2}$.
It shows that from the viewpoint of the observer 
at infinity a particle dropped from rest begins to accelerate, then at some point  
starts decelerating and eventually stops at $r=2r_g$. The bigger initial kinetic 
energy, the greater a "resisting" force arising so that the speed of the particle cannot exceed the 
coordinate speed of light. Strangely enough, if $\gamma_0 \ge\sqrt{3/2}$, 
the particle will never accelerate in a gravitational field, (see (\cite{Okun, Mashhoon}, and elsewhere). 

The GR formula should be compared with our result (\ref{39}):\  \
 $\beta(r)=\left[1-(1/\gamma_0^2)\exp(-2r_g/r)\right]^{1/2}$, which does not indicate any ``resisting force''.
 


\medskip
{\it 6. ``Black holes'' and other ``strong field'' observations}

There are astrophysical observations related to strong-field effects (the so-called black holes, radiating binary star systems, and others). Of course, there should be strong-field effects around astrophysical objects of super-high density. Practically, they might look like circumstantial evidence of ``black holes'' manifesting ``gravitational collapse'' and the corresponding ``light trap''. However, the idea of matter collapse into a singularity point in space seems to be an unnecessary ``new physics'' speculation. In our alternative approach, the gravitational time dilation could be however great; physical processes involving particle and photon motion in a strong field remain time-reversal and free of singularities. We predict an existence of extremely dense ordinary material formations of a strong gravitational pull without collapsing.

\section{Coulomb Force}

\subsection{Equation of radial motion}

We are not interested in effects due to a magnetic field; our topic is narrowed to an energy balance for an electron in the Coulomb field. A classical attractive electric (Coulomb) force on an electron of a charge $e$ and a proper mass $m_0$ due to a source of a charge $Q>>e$ and a mass $M>>m$ in the approach of variable proper mass is
\begin{equation}
F_e=\frac{kQe}{r^2}
\label{53}
\end{equation}
where $k$ is the electric constant, and the electric radius $r_e=keQ/m_0c_0^2$ is analogous to the gravitational radius $r_g=GMm/m_0c_0^2$. The $r_e$ is an electric field strength parameter giving a criterion of weak-field conditions $r_e/r<<1$ for the electron. 

Bearing in mind the total energy conservation $\gamma m=m_0$, $\gamma=\gamma_0 \gamma_r$ in a full analogy to the gravitational force, it follows for a free radial fall from rest at infinity
\begin{equation}
F_edr=m_0c_0^2 d(r_e/r)=m_0 v dv,  \ \ \ \beta d\beta=d(r_e/r)  
\label{54}
\end{equation}
with the solution
\begin{equation}
m/m_0=1/\gamma_r=(1-2r_e/r)^{1/2}, \  \  (r\ge R\ge r_e)
\label{55}
\end{equation}

It is assumed further that the electric field does not affect a gravitational interaction, and a ratio of gravitational to Coulomb force does not depend on a gravitational or electric field strength. Hence, $k$ should differ from $k_0$ at infinity; namely
\begin{equation}
k(r)/k_0=1/{\gamma_r}^2
\label{56}
\end{equation}
Then the Coulomb force has exactly the same form as the gravitational force: all formulae obtained for the gravitational force are valid for the Coulomb force after replacing $r_g$ by $r_e$. In particular, we have
\begin{equation}
1/\gamma_r=m(r)/m_0=\exp(-r_e/r), \ \ r\ge R
\label{57}
\end{equation}
\begin{equation}
m/m_0=1/\gamma_r=\exp(-r_e/r), \  \  (r\ge R)
\label{58}
\end{equation}
\begin{equation}
\gamma=\gamma_0\exp(r_e/r), \ \ \beta=\left[1-(1/\gamma_0)\exp(-2r_e/r)\right]^{1/2} 
\label{59}
\end{equation}
\begin{equation}
p=c_0 \gamma_0 \gamma \beta m=c_0 \gamma_0 \beta m_0
\label{60}
\end{equation}
and the expressions for total, kinetic, and potential energy:
\begin{equation}
E_{tot}=\gamma_0^2 m_0^2 c_0^4=p^2c_0^2+m_0^2 c_0^4 
\label{61}
\end{equation}
\begin{equation}
E_{kin}=E_{tot}-mc_0^2=m_0 c_0^2 \left[\gamma_0-\exp(-r_e/r)\right] 
\label{62}
\end{equation}
\begin{eqnarray}
W(r)=-m_0 c_0^2\left[ 1-\exp{(-r_e/r)} \right] \ , &   (r\ge R)
\label{63} 
\end{eqnarray}

\subsection{A charged particle in the Coulomb field: a test proposal} 

In the conventional Relativistic Dynamics \cite{Landau} the equation of relativistic motion of a point charge in the Coulomb field is 
\begin {equation}
m_0d(\gamma v^i)/dt=F_e
\label{64}
\end{equation} 
where the proper mass of the electron $m=m_0$ and the electric constant 
$k=k_0$ are field independent physical constants. Thus, the conventional equation of motion is
\begin{equation}
m_0 \frac{d}{dt}(\gamma v)=-\frac {k_0Qe}{r^2}dr, \ \ (r\ge R)
\label{65}
\end{equation}
or equivalently 
\begin{eqnarray}
\beta d(\gamma\beta)=d(r_e/r),  \   d\gamma=d(r_e/r), \  \   (r\ge R)  
\label{66}
\end{eqnarray}
with the solution for free fall from rest at infinity ($\gamma_0=1$)
\begin{equation}
\gamma=1+r_e/r, \ \  \beta=\sqrt{1-(1+r_e/r)^{-2}} 
\label{67}
\end{equation}
\begin{equation}
p=\gamma\beta m_0c_0=\sqrt{(1+r_e/r)^2-1]}   
\label{68}
\end{equation}
\begin{equation}
E_{tot}=m_0c_0^2(1+r_e/r)=\gamma m_0c_0^2  
\label{69}
\end{equation}
\begin{equation}
E_{kin}=(\gamma-1)m_0c_0^2=m_0c_0^2(r_e/r)=k_0Qe/r  
\label{70}
\end{equation}
or for $r=R$
\begin{equation}
E_{kin}=k_0Qe/r=eV_{s} \  \  (r\ge R)  
\label{71}
\end{equation}
where $V_{s}$ is a positive voltage of the attractive spherical shell. Clearly, from (\ref{71})  it follows that the kinetic energy of the electron falling onto the charged shell indefinitely rises with the potential. It should be compared with the alternative prediction (\ref{62})
\begin{equation}
E_{kin}=m_0c_0^2(1-\exp (-eV_{s}/m_0c_0^2) \  \  (r\ge R)  
\label{72}
\end{equation}
which reads that the electron cannot be accelerated to energies higher than $m_0c_0^2$=511 keV regardless of how great the potential (the voltage) of an electrostatic spherical conductor is. In fact, it means that in a relativistic theory the unit of $eV$ needs to be strictly defined due to non-linearity under strong field conditions (what is the case for the electron in a high potential (strong) field according to the criterion 
$eV_{s}>>m_0 c_0^2= 511 \ keV$). It should be noted that for a proton the similar conditions occur at much higher potential (by factor $m_p /m_e \approx 1833$). In the conventional Electrodynamics, an electron kinetic energy (\ref{71}) obeys a linear law, which is the case in the alternative theory under weak-field conditions. Our statement is that the conventional theory is, in fact, a linear approximation of a more general (non-linear) SR-based particle Dynamics theory.

Measurements of energy of electrons accelerated by the attractive Coulomb field can be conducted at Van de Graaf (static accelerator) facilities. Typical construction of the accelerator includes a source of particles, a spherical shell (a conductor) isolated from ground, and a focusing system. The conductor may be ``pumped'' by an electric transporter up to millions of volts of a positive or negative potential. In a usual regime, the machine accelerates beams of protons and other positively charged particles up to the energy of a positive potential (in electron Volt units). Sometime, electrons accelerated in the repulsive potential field due to a negatively charged conductor are used. In this case, the sign of $r_e$ in (\ref{63}) changes. Consequently, the initial proper mass of accelerated electrons would be bigger than $m_0$: it exponentially grows with the potential. The difference will contribute to kinetic energy, and non-linear effect can be measured again. To our knowledge, there is no systematical empirical data available concerning the above suggested tests with electrons in attractive or repulsive high potential Coulomb field.

\section {Summary and Conclusion}

-  The problem of relativistic motion in a gravitational field is studied in the Special Relativity dynamics of point particle. A novelty of our approach to the problem is the introduction of the field dependent proper mass concept, as opposed to the conventional assumption of proper mass constancy. Historically, the SR-based gravitational dynamics has never been developed. It was believed that the gravity phenomenon and Special Relativity are incompatible, mostly because of the difficulties of a photon property description. General Relativity did explain the observed gravitational properties of particles and photons. As of today, GR has been thoroughly tested under weak-field conditions; however, strong-field effects still have not been verified in direct measurements. The long-standing, not thoroughly understood problem is the GR non-quantizibility. Another problem is associated with the strong-field $1/r$ divergence, which cannot be removed by a means of renormalization procedure. That is why alternative approaches to the gravitational problem are often speculated in literature.

-  We studied conservation properties of the $1/r$ gravitational potential in the relativistic Lagrange framework in the context of Noether's currents associated with the time and mass translation symmetry. The complementarity principle in relativistic dynamics is formulated, and quantum connections of the theory via the generalized de Broglie wave concept are established taking into account that the proper mass and time are scalars, which determine the temporal part of coordinate and momentum (complementary) 4-vectors characterizing the particle as a standard quantum oscillator, or a standard atomic clock. The complementarity principle requires that all observers use standard atomic clocks in the metric determination (time pace) in the field as compared to the time pace at infinity. The principle enables us to gain an insight into a unity of mass and time concepts and quantum  connections of relativistic gravitational dynamics due to the relationship $m_0 c_0^2 \Delta\tau_0= h$.

-  One of our findings is that a photon propagation may be described in terms of refraction in a gravitational field medium. This is not an unusual approach to the photon problem: a similar photon concept is from time to time considered in the GR. It means that the photon may not be energetically coupled to the gravitational field but be influenced by another (refraction) mechanism of gravitational interaction. We conclude that the inclusion of the photon refraction concept along with a revised proper mass concept into SR-based mechanics makes predictions consistent with existing gravitational (weak-field) observations. New predictions in strong field domain are made.

-  The photon does not have the proper mass. Consequently, it gives rise to the null Lorentzian metric. In SR methodology, the photon plays an important role in determination of both temporal and spatial parts of complementary 4-vectors by a means of information (photon) exchange between observers. As a result of our study, we conclude that the relativistic Lagrangean problem has a dual formulation in terms of complementary quantities. This makes the concept of a gravitational field as a refracting medium more understandable but still does not give a clue about the mechanism of changing the speed of light (permittivity and permeability of space) in the field. The question of ``refracting'' properties of a gravitational field must be challenging for quantum gravity researchers.

-  Our conclusion that the source of the gravitational and Coulomb potential energy is the proper mass needs to be further studied, in particularly, a prediction that kinetic energy of a point charge in a microscopic Coulomb field is due to a change of the proper mass of a charged particle (similarly to the case of a point particle in the gravitational field due to the mass source). This prediction can be tested in experiments with electrons in an electrostatic laboratory.

-  One of the motivations of this work is new predictions in a strong field domain, in particularly, the $1/r$ divergence elimination through a natural mechanism of mass defect rising with field strength (the ``mass exhaustion'' effect).

-  We believe that the variable proper mass approach developed in the SR-based mechanics framework will be useful for better understanding of the GR problems and also perspective for further studies on developing divergence-free field theories.

\end{document}